\pageno=0
\footline={\ifnum\pageno=0\hfill\else\tenrm\hfill
\folio\hfill\fi}
\magnification\magstep 1
\baselineskip 18pt
 3

\font\tme=cmr8
\font\eightit=cmti8
\newcount\eqnno\eqnno=0
\def\nexteqn{\global\advance\eqnno by 1 (\the\eqnno)}
\def\num{\eqno\nexteqn} \rm
\def\nabl{\nabla\!}
\null
{\baselineskip=10pt
\centerline {\bf 3+1 Approach to the Long Wavelength Iteration Scheme}}
\vskip 1.5cm {\baselineskip=10pt
\centerline {G. L. Comer${}^{*}$}
\centerline {\eightit Department of Science and Mathematics}
\centerline {\eightit Parks College of Saint Louis University}
\centerline {\eightit Cahokia, IL 62206, USA}}
\vskip 1cm
\centerline {October 30, 1996}
\vskip 1cm
{\tme\baselineskip=10pt\centerline {ABSTRACT}}
Large-scale inhomogeneities and anisotropies are modeled using the Long
Wavelength Iteration Scheme.  In this scheme solutions are obtained as
expansions in spatial gradients, which are taken to be small.  It is 
shown that the choice of foliation for spacetime can make the iteration 
scheme more effective in two respects: (i) the shift vector can be chosen 
so as to dilute the effect of anisotropy on the late-time value of the 
extrinsic curvature of the spacelike hypersurfaces of the foliation; and 
(ii) pure gauge solutions present in a similar calculation using the 
synchronous gauge vanish when the spacelike hypersurfaces have extrinsic 
curvature with constant trace.  We furthermore verify the main conclusion 
of the synchronous gauge calculation which is large-scale inhomogeneity 
decays if the matter---considered to be that of a perfect-fluid with a 
barotropic equation of state---violates the strong-energy condition.  
Finally, we obtain the solution for the lapse function and discuss its 
late-time behaviour.  It is found that the lapse function is well-behaved 
when the matter violates the strong energy condition.      
\bigskip
\noindent
${}^{*}$E-mail: comer@newton.slu.edu
\vfill
\eject
\noindent{{\bf 1. Introduction}}
\bigskip
The long wavelength iteration scheme [1,2,3] has been used to model 
analytically cosmologies with large scale inhomogeneity.  This scheme was 
first introduced by Lifschitz and Khalatnikov (see [4] and references 
therein).  A similar scheme has been used by Salopek, Stewart, and 
collaborators [5] in their analysis of the Hamilton-Jacobi equation for 
general relativity.  It has also been applied to Brans-Dicke theory [6] 
and the more general scalar-tensor gravity [7].  The essential part of the 
long wavelength iteration scheme is spatial gradients are considered to be 
small.  Therefore, the scheme immediately suffers from the fact that there 
is no absolute space, and hence no absolute notion of spatial gradient, 
large or small.  We circumvent this difficulty by using the 3+1 formalism 
to perform our calculations (see York [8] for an excellent review).  The 
technical problems associated with the lack of an absolute space are 
overcome by introducing a foliation of spacetime, and hence a notion of 
time and the flow of time.  But furthermore, we will see that the 3+1 
formalism can also make the long wavelength iteration scheme more 
effective at producing inhomogeneous and anisotropic cosmological models.  

Refs. [1,2,3] solve a first-order ({\it i.e.}, one time derivative) form 
of the field equations where the synchronous gauge is imposed.  In the 
language of the 3+1 formalism, the synchronous gauge sets the shift vector 
to zero and the lapse function to one.  But the immediate simplification 
this gauge choice obtains, in making the field equations less complicated 
by having fewer fields to worry about, does not necessarily make the 
calculations less complicated (see, for instance, Piran [9], Smarr and 
York [10], or Hawking and Ellis [11]).  For instance, it will be shown 
below that the shift vector can be used to exponentially damp (in time) 
the influence of anisotropy on the value of the extrinsic curvature of the 
spacelike leaves of the foliation.  This can be significant because in the 
synchronous gauge approach obtaining the three-metric of the spacelike 
slices of the foliation when anisotropy is present is complicated and 
was only treated as a linear perturbation during the iterative part of 
the calculations.

The most important difference between the formulation used here and that 
of Ref. [1] is in the choice of lapse function: we will use a lapse 
function such that the trace of the extrinsic curvature, $K$, of each 
spacelike slice associated with the foliation is constant.  The value of 
$K$ will change from one slice to the next, and is thus a function of 
time, {\it i.e.}, $K = k(t)$. (The trace can even be taken to be the time 
itself [12], which is known as `York time.')  The $K = k(t)$ slicing 
condition is used because it ensures better than the synchronous gauge 
that calculated inhomogeneity is physical [12,13,14], and not merely an 
artifact of embedding unnecessarily distorted spacelike slices into an 
otherwise homogeneous spacetime.  Indeed, it will be seen below that pure 
gauge solutions appearing in the results of Comer {\it et al.} [1] do not 
show up here.  However, we do find their other solutions and thus confirm 
the physical origin of the inhomogeneity they calculate.  Furthermore, we 
also confirm their main conclusion, which is long wavelength inhomogeneity 
grows or decays, depending on the choice of the equation of state for the 
matter.  In particular, the inhomogeneity decays when the matter is 
`inflationary' ({\it i.e.}, it violates the strong energy condition [11]).
      
It must be stated now that we do not solve the 3+1 form of the field 
equations as an initial-value problem.  That is, no explicit attempt is 
made to solve first the constraints---so as to isolate the true degrees of 
freedom, whose initial values make up the initial data set---and then use 
the remaining field equations to evolve the initial data set forward in 
time.  Here expansions in terms of spatial gradients of a ``seed'' metric 
are made for the three-metric of the spacelike slices and matter variables.  
The zeroth-order term in an expansion contains no spatial gradients, the 
second-order term, which is next, contains two spatial derivatives, the 
fourth-order has four, and so on.  (We shall demonstrate that the lapse 
function, but not the shift vector, must also be expanded in this way.)  
The coefficients in the expansions are constructed so that the 
constraints, as well as the other equations, are satisfied order-by-order.  
In this context, the initial-value problem becomes just a different way of 
solving the field equations; however, we will discuss it briefly in the 
concluding section. 

Building a slicing of spacetime based on keeping the trace of the 
extrinsic curvature constant has now a long history, beginning with the 
maximal slicing condition of Lichnerowicz [15].  Since then much work has 
been done to determine when the $K = k(t)$ condition may be used, for 
instance, in the context of homogeneous cosmology (see Ryan and Shepley 
[16] for a review) or vacuum,  asymptotically flat spacetimes with no 
restriction to homogeneity [17].  Goddard [18] has even shown for closed 
universes that if one $K = k(t)$ slice exists, and the strong energy 
condition is satisfied, then the $K = k(t)$ condition may be applied 
through the whole spacetime.  More recently, Goldwirth and Piran [19] 
demonstrated numerically that the $K = k(t)$ slicing condition breaks down 
for closed, inhomogeneous universes having `inflationary' matter.  What 
they find is one cannot prescribe a $k(t)$ which is a monotonic function 
of time and simultaneously construct an acceptable lapse function.  
However, they were able to use $K = k(t)$ for open universes, even during 
inflation, with no problems.  As for the present work, there are no 
difficulties with using the $K = k(t)$ slicing condition when the matter 
is `inflationary.'  We can specify a monotonic $k(t)$ and still get an 
acceptable solution for the lapse function (in the sense that it remains 
positive for all time).  On the other hand, when the matter satisfies the 
strong energy condition, caution must be exercised in the use of $K = 
k(t)$, especially when the higher-order terms in the lapse function become 
comparable to the zeroth-order term.  
  
In Sec. 2 we give the 3+1 form of the field equations and set the bulk 
of the notation.  In Sec. 3 we discuss in general terms ({\it i.e.}, for 
arbitrary lapse function and shift vector) the long wavelength 
approximation [20].  In Sec. 4, we discuss the zeroth-, second-, and 
fourth-order equations and solutions for $K = k(t)$ hypersurfaces.  In 
Sec. 5, we write the solution for the lapse function, good to fourth-order 
in spatial gradients, and discuss how it behaves at late times.  In Sec. 
6, we offer some concluding remarks.  Finally, some formulas used to 
construct the second- and fourth-order equations are listed in Appendix A 
and the method through which the zeroth-order solution for the extrinsic 
curvature is obtained is given in Appendix B.  We will use units 
throughout such that $G = c = 1$ as well as `MTW' conventions (see Ref. 
[21]).
\bigskip
\noindent{{\bf 2. The 3+1 Decomposition and Field Equations}}
\bigskip
The notion of `time' in the 3+1 formalism enters via a scalar function 
$t = t(x^{\mu}),~\mu = 0,1,2,3,$ whose level surfaces are spacelike.  The 
flow of time is introduced via a vector $t^{\mu}$, which can be either 
timelike or spacelike.  The vector which is perpendicular to the level 
surfaces of $t(x^{\mu})$ is $n_{\mu} = - N \nabl_{\mu} t$, where the lapse 
function $N$ ensures that $n^{\nu} n_{\nu} = - 1$.  The vector $t^{\mu}$ 
is not parallel to $n^{\mu}$ but has the form 
$$t^{\nu} = \left(\perp^{\nu}_{\mu} -n^{\nu} n_{\mu}\right) t^{\mu} 
\equiv N^{\nu} + N n^{\nu} \ , \num$$
where $\perp^{\nu}_{\mu} = \delta^{\nu}_{\mu} + n^{\nu} n_{\mu}$ is the
`projection-operator' that projects spacetime tensors into the level 
surfaces obtained from $t(x^{\mu})$ and $N^{\mu}$ is the shift vector 
($n_{\nu} N^{\nu} = 0$). From Eq. (1), it is seen that $N$ must always 
remain positive, otherwise there would exist spacetime points where 
$t^{\mu}$ would be parallel to the spacelike leaves of the foliation [10].  
It must also be emphasized that the lapse function and shift vector are 
arbitrary in the sense that the four independent functions associated with 
them are not determined by the field equations but instead must be 
specified by hand.     

From here on out it will be assumed that a local coordinate system
$x^{\mu} = (x^0, x^i)$ exists, where the $x^i,~i = 1,2,3,$ are the 
coordinates of the spacelike slices associated with the foliation and we 
take $t = x^0$.  With respect to these coordinates $N^{\mu} = (0,N^i)$, 
and the line interval can be written
$${\rm d}s^2 = - (N^2 - N^i N_i) {\rm d}t^2 + 2 N_i {\rm d} x^i {\rm d} t 
+ \gamma_{i j} {\rm d} x^i {\rm d} x^j \ , \num$$
where $\gamma_{i j} = g_{i j}$ and $N_i = \gamma_{i j} N^j$.  The spatial
covariant derivative compatible with $\gamma_{i j}$ is denoted 
$\tilde{D}_i$.

The 3+1 field equations consist of four constraint equations and twelve
evolution equations (see Ref. [8] for more discussion).  They contain 
terms with only one time derivative because $\dot{\gamma}_{i j}$ (where a 
dot overscript `$\dot{~}$' means $\partial/\partial t$) is replaced by the
extrinsic curvature, whose spatial components $K_{i j}$ are obtained from 
[22]
$$K_{i j} = - {1 \over 2}~\perp^{\sigma}_{i} \perp^{\tau}_{j}
\nabl_{\sigma} n_{\tau} = {1 \over 2}~N^{- 1} \left[\tilde{D}_i N_{j} + 
\tilde{D}_j N_{i} - \dot{\gamma}_{i j}\right] \ . \num$$
The inverted form of Eq. (3) thus represents six of the evolution 
equations (for $\gamma_{i j}$).  The other evolution equations, and the 
constraints, result from projecting the free indices of the Einstein 
equations into and out of the spacelike slices of the foliation.  The 
three independent projections of the stress-energy tensor $T_{\mu \nu}$ 
define an energy density $\rho \equiv n^{\mu} n^{\nu} T_{\mu \nu}$, 
a three-current ${\cal J}_i \equiv - n^{\nu} \perp^{\sigma}_{i} T_{\nu 
\sigma}$, and spatial stress tensor ${\cal S}_{i j} \equiv 
\perp^{\sigma}_{i} \perp^{\tau}_{j} T_{\sigma \tau}$.  The 3+1 equations 
are 
$$16 \pi \rho = \tilde{R} - K^i_j K^j_i + K^2 \ , \num$$
$$8 \pi {\cal J}_i = \tilde{D}_j \left(K^j_i - K \delta^j_i\right) \ , 
\num$$
$$8 \pi {\cal S}_i^j = \tilde{R}_i^j + N^{- 1} \left(K_k^j \tilde{D}_i N^k 
- K^k_i \tilde{D}_k N^j\right) + K K_i^j -  N^{- 1} \dot{K}_i^j -$$ 
$$N^{- 1} \tilde{D}_i \tilde{D}^j N + N^{- 1} N^k \tilde{D}_k K_i^j - 4 
\pi (\rho - {\cal S}) \delta_i^j \ , \num$$
and
$$\dot{\gamma}_{i j} = \tilde{D}_i N_j + \tilde{D}_j N_i - 2 N K_{i j} 
\ , \num$$
where $\tilde{R}_{i j}$ is the three-dimensional Ricci tensor formed from 
$\gamma_{i j}$, $\tilde{R} = \gamma^{i j} \tilde{R}_{i j}$ is its 
associated Ricci scalar, and ${\cal S} \equiv \gamma^{i j} 
{\cal S}_{i j}$.  Note that Eq. (6) is the trace free form of 
$\perp^{\sigma}_{i} \perp^{\tau}_{j} G_{\sigma \tau} = 8 
\pi{\cal S}_{i j}$, the removed trace being
$$\dot{K} = 4 \pi N \left(\rho + {\cal S}\right) + N^i \tilde{D}_i K + 
N K^j_i K^i_j - \tilde{D}^i \tilde{D}_i N \ . \num$$
Lastly, the evolution equation for the determinant of the three-metric 
$\gamma$ (which will be used below) is
$$\dot{\gamma} = 2 \gamma \left(\tilde{D}_i N^i - N K\right) \ . \num$$

The field equations for a perfect fluid are $\nabla_{\nu} T^{\nu}_{\mu} = 
0$.  The independent projections of the free index gives
$$\dot{\rho} + N \tilde{D}_i {\cal J}^i = N \left(K^i_j {\cal S}^j_i + K
\rho\right) - 2 {\cal J}^i \tilde{D}_i N + N^i \tilde{D}_i \rho \num$$
and
$$\dot{{\cal J}}^i + N \tilde{D}_j {\cal S}^{i j} = N \left(2 K^i_j 
{\cal J}^j + K {\cal J}^i\right) - {\cal S}^{i j} \tilde{D}_j N - \rho 
\tilde{D}^i N + N^j \tilde{D}_j {\cal J}^i - {\cal J}^j \tilde{D}_j N^i 
\ , \num$$
where $\rho$, ${\cal J}^i$ and ${\cal S}_{i j}$ are obtained from the 
projections of the perfect fluid stress-energy tensor
$$T_{\mu \nu} = \left(\rho^* + p^*\right) u_{\mu} u_{\nu} + p^* 
g_{\mu \nu} \ . \num$$
The fluid four-velocity $u^{\mu}$ has the normalization $u^{\mu} u_{\mu} 
= - 1$ and the pressure $p^*$ is related to the energy density $\rho^*$ 
via the equation of state $p^* = \left(\Gamma - 1\right) \rho^*$, where 
$0 \leq \Gamma \leq 2$.  Letting $\tilde{u}^i$ represent the projection 
of the four-velocity into the slices of the foliation, {\it i.e.}, 
$\tilde{u}^i = \perp^i_{\sigma} u^{\sigma}$, and $|{\bf \tilde{u}}|^2 
\equiv \tilde{u}^{i}~\tilde{u}_{i}$, then
$$\rho = \rho^* \left(1 + \Gamma |{\bf \tilde{u}}|^2\right) \ , \num$$
$${\cal J}_i = \pm \Gamma \rho^* \sqrt{1 + |{\bf \tilde{u}}|^2}~
\tilde{u}_i \ , \num$$
and
$${\cal S}_{i j} = \Gamma \rho^*~\tilde{u}_i~\tilde{u}_j + \left(\Gamma 
- 1\right) \rho^* \gamma_{i j} \ . \num$$
\bigskip
\noindent{{\bf 3. The Long Wavelength Approximation}}
\bigskip
In this section we invoke the long wavelength approximation, where spatial
gradients are neglected entirely.  These solutions are accurate 
representations of cosmologies wherein deviation from perfect homogeneity
occurs on scales bigger than the co-moving Hubble length (see Tomita [20], 
and also Ref. [1], for more discussion).  It will be seen that one can 
extract a significant amount of information about the solutions before 
any assumptions about the lapse function are made and without setting the 
shift vector to zero.  In particular, since the lapse function and shift 
vector are specified freely, there is no a priori reason for their spatial 
gradients to vanish.  

The long wavelength equations are
$$16 \pi \rho_{0} \approx - {}^{(0)}K^i_j~{}^{(0)}K^j_i + {}^{(0)}K^2 \ ,
\num$$
$$8 \pi~{}^{(0)}{\cal J}_i \approx 0 \ , \num$$
and
$$8 \pi {}^{(0)}{\cal S}_i^j \approx N^{- 1} \left({}^{(0)}K_k^j 
\tilde{D}_i N^k - {}^{(0)}K^k_i \tilde{D}_k N^j\right) + {}^{(0)}K~
{}^{(0)}K_i^j - N^{- 1}~ {}^{(0)}\dot{K}_i^j -$$ 
$$N^{- 1} \tilde{D}_i \tilde{D}^j N - 4 \pi (\rho_0 - {}^{(0)}{\cal S})
\delta_i^j \ , \num$$
where a `$0$' left-superscript (or right-subscript) on any quantity 
signifies it is zeroth-order in spatial gradients.  The equations for 
$\dot{\gamma}_{i j}$ and $\dot{\gamma}$ do not change their form from the 
relations given in the last section; however, that for $\dot{K}$ does 
change to
$${}^{(0)}\dot{K} \approx 4 \pi N \left(\rho_0 + {}^{(0)}{\cal S}\right) 
+ N~{}^{(0)}K^j_i~{}^{(0)}K^i_j - \tilde{D}^i \tilde{D}_i N \ . \num$$
The stress-energy divergence equations become, by forcing Eq. (17) to 
hold on each spacelike slice (which means taking ${}^{(0)}\tilde{u}_{i} 
\approx 0$ at all times),
$$\dot{\rho}_0 \approx N \left({}^{(0)}K^i_j~{}^{(0)}{\cal S}^j_i +
{}^{(0)}K~\rho_0\right) \num$$
and
$$0 \approx \left({}^{(0)}{\cal S}^{j}_{i} + \rho_0 \delta^{j}_{i}\right)
\tilde{D}_j N \ . \num$$

Two immediate conclusions follow from Eqs. (20) and (21): using the 
evolution equation for $\gamma$, then it can be obtained from Eq. (20) 
that
$$\rho_0~{}^{(0)}\gamma^{\Gamma/2} \approx {\cal C}~{\rm exp} \left(\Gamma 
\int_{t_0}^t {\rm d} \tilde{t}~\tilde{D}_i N^i\right) \ , \num$$
where ${\cal C}$ depends only on $x^i$.  Eq. (21) says that $\tilde{D}_i 
N \approx 0$ whenever the combination ${}^{(0)}{\cal S}^{j}_{i} + \rho_0
\delta^{j}_{i}$ is not zero.  (For a perfect fluid with the equation of 
state used here, the combination is zero only when $\Gamma = 0$.)  Hence, 
even though the lapse function is in principle freely specifiable, 
consistency of the long wavelength approximation demands that it have a 
small spatial gradient.  

The remaining equations (18) and (7) are to be solved, respectively, to 
obtain ${}^{(0)}K^j_i$ and then ${}^{(0)}\gamma_{i j}$.  Finding 
${}^{(0)}K^j_i$ requires a rewriting of Eq. (18).  This is accomplished 
by using Eq. (16), and the fact that $\tilde{D}_i N$ and the three-velocity 
are both zero at this order.  Also letting ${}^{(0)}{\cal K}^j_i \equiv 
\sqrt{{}^{(0)}\gamma}\left[{}^{(0)}K^j_i - {1 \over 3}~{}^{(0)}K 
\delta^j_i\right]$, which is the trace-free part of ${}^{(0)}K^j_i$ 
(modulo the $\sqrt{{}^{(0)}\gamma}$ factor), then Eq. (18) becomes 
$${}^{(0)}\dot{\cal K}^j_i - {\cal L}_{\bf N} {}^{(0)}{\cal K}^j_i \approx 0 
\ , \num$$
where ${\cal L}_{\bf N} {}^{(0)}{\cal K}^j_i$ is the Lie derivative with
respect to $N^i$, {\it i.e.}, ignoring spatial gradients of the extrinsic
curvature and the three-metric,
$${\cal L}_{\bf N} {}^{(0)}{\cal K}^j_i \approx \left(\tilde{D}_k 
N^k\right) {}^{(0)}{\cal K}^j_i + \left(\delta^j_l \tilde{D}_i N^k - 
\delta^k_i \tilde{D}_l N^j\right) {}^{(0)}{\cal K}^l_k \ . \num$$
In order to solve Eq. (23), it is convenient to interpret it as a linear 
vector equation (with the vector ${\cal \bf K}$ being formed out of the 
components of ${}^{(0)}{\cal K}^j_i$), {\it i.e.},
$$\dot{\cal \bf K} - \left(\tilde{D}_k N^k\right) {\cal \bf K} - {\cal 
\bf M} {\cal \bf K} \approx 0 \ , \num$$
where 
$${\cal \bf K} = [{}^{(0)}{\cal K}^1_1~{}^{(0)}{\cal K}^1_2~{}^{(0)}
{\cal K}^1_3~{}^{(0)}{\cal K}^2_1~{}^{(0)}{\cal K}^2_2~{}^{(0)}{\cal 
K}^2_3~{}^{(0)}{\cal K}^3_2~{}^{(0)}{\cal K}^3_2~{}^{(0)}{\cal 
K}^3_3]^{\rm T} \num$$
and
$${\cal \bf M} = \left[\matrix{0&a_4&a_7&-a_2&0&0&-a_3&0&0 \cr
a_2&a_5-a_1&a_8&0&-a_2&0&0&-a_3&0 \cr a_3&a_6&a_9-a_1&0&0&-a_2&0&0&-a_3 
\cr -a_4&0&0&a_1-a_5&a_4&a_7&-a_6&0&0 \cr 0&-a_4&0&a_2&0&a_8&0&-a_6&0 
\cr 0&0&-a_4&a_3&a_6&a_9-a_5&0&0&-a_6 \cr -a_7&0&0&-a_8&0&0&a_1-a_9&a_4&
a_7 \cr 0&-a_7&0&0&-a_8&0&a_2&a_5-a_9&a_8 \cr 0&0&-a_7&0&0&-a_8&a_3&a_6&0}
\right]\ , \num$$
with $a_1 = \tilde{D}_1 N^1$, $a_2 = \tilde{D}_2 N^1$, $a_3 = 
\tilde{D}_3 N^1$, $a_4 = \tilde{D}_1 N^2$, $a_5 = \tilde{D}_2 N^2$, 
$a_6 = \tilde{D}_3 N^2$, $a_7 = \tilde{D}_1 N^3$, $a_8 = \tilde{D}_2 N^3$, 
and $a_9 = \tilde{D}_3 N^3$.  One can quickly verify that ${\rm Tr} 
{\cal \bf M} = 0$.  Therefore, the eigenvalues of ${\bf M}$ must add up 
to zero.

Even though Eq. (25) is linear in the extrinsic curvature, solving it for 
a general shift vector is a daunting linear algebra eigenvalue problem, 
further complicated by the time dependence in the matrix ${\cal \bf M}$.  
We will simplify the problem by using the freedom to specify, by hand, the 
shift vector and assume that its spatial gradient is time-independent.  
Nevertheless, finding the eigenvalues of the matrix ${\cal \bf M}$ still 
means solving a ninth-order polynomial equation.  Fortunately, a lot of 
progress can be made even without knowing explicitly the eigenvalues.  By 
following the procedure outlined in Appendix B, then it can be shown that 
$${}^{(0)}{\cal K}^j_i = {\rm exp}{\left[\left(\tilde{D}_k N^k\right) t
\right]} {\cal G}^j_i \ . \num$$
The tensor ${\cal G}^j_i(x^k,t)$ is the analog of the so-called anisotropy
matrix obtained in the synchronous gauge calculations of Ref. [1].  The 
big difference between our result and that of Ref. [1] is the anisotropy 
matrix of Ref. [1] is independent of time whereas ${\cal G}^j_i$ has an 
exponential time dependence.  (Clearly, ${\cal G}^j_i$ becomes 
time-independent when the shift vector is zero.)  Hence, we see explicitly 
that one can artificially remove or enhance the anisotropy coming from 
${\cal G}^j_i$, depending on what choice is made for the shift vector.  
More to the point, let us consider the simplified example where only 
$a_1$, $a_5$, and $a_9$ are non-zero, with $a_1 = a_5 = a_9 < 0$.  Then 
it is easy to see that ${}^{(0)}{\cal K}^j_i$ decays to zero exponentially
with time.  Hence, as $t$ increases, it will quickly be the case that
$${}^{(0)}K^j_i = {1 \over 3}~{}^{(0)}K \delta^j_i \ . \num$$

The last item for this section is to determine ${}^{(0)}\gamma_{i j}$.  
Since we have seen how to use the shift vector to reduce the effect of 
${\cal G}^j_i(x^k,t)$ on ${}^{(0)}{\cal K}^j_i$, we will henceforth set 
both $N^i$ and ${\cal G}^j_i$ to zero for the rest of this work.  Eq. (29) 
thus becomes the zeroth-order solution for the extrinsic curvature.  From 
Eqs. (7), (9), and (29) it follows that 
$${}^{(0)}\gamma_{i j} = {}^{(0)}\gamma^{1/3} h_{i j} \qquad , \qquad 
h^{j k} h_{k i} = \delta^j_i \ , \num$$
where $h_{i j}$ depends only on $x^i$ and is the so-called ``seed'' metric 
mentioned in the introduction.  The remaining field equations only give 
one more thing, when Eq. (22) is used, and that is
$${}^{(0)}K = - \sqrt{24 \pi {\cal C}}~~{}^{(0)}\gamma^{- \Gamma/4} \ . 
\num$$
To determine the explicit time-dependence of ${}^{(0)}\gamma$, and thus 
the other zeroth-order quantities, the lapse function must be specified.  
For instance if $N = 1$, like in the synchronous gauge, then ${}^{(0)}
\gamma^{1/3} \propto t^{4 \Gamma/3}$.  Or, a condition like $K = k(t)$ 
must be imposed, which is what we do next.
\bigskip
\noindent{{\bf 4. The Long Wavelength Iteration Scheme}}
\bigskip
The long wavelength iteration scheme obtains solutions to the Einstein 
equations by assuming spatial gradients of the gravitational and matter 
fields are small on every spacelike slice of the foliation.  The solutions 
contain terms with ever-increasing spatial gradients of the ``seed'' 
metric obtained in the last section.  To have solutions with inhomogeneity 
on shorter and shorter scales, terms with more and more spatial gradients 
must be included [23].  Here we write down solutions with terms containing 
up to four gradients.  

To be precise, the field variables $\rho$, $\tilde{u}^i$, and 
$\gamma_{i j}$ are expanded as follows (letting $A_0(t) 
\equiv {}^{(0)}\gamma^{1/3}$):
$$\rho = \rho_0(t) + \rho_{\rm 2}(t) R + \rho_4 R^2 + \mu_4(t) R_{i j} 
R^{i j} + \epsilon_4 D_i D^i R + ... \ , \num$$
$$\tilde{u}_i = u_3(t) D_i R + u_5(t)R D_i R + v_5(t) R^j_i D_j R + w_5(t) 
R^j_k D_i R_j^k +$$
$$x_5(t) R^k_j D_k R^j_i + y_5(t) D_i D_j D^j R ... \ , \num$$
and
$$\gamma_{i j} = {}^{(0)}\gamma_{i j} + {}^{(2)}\gamma_{i j} + 
{}^{(4)}\gamma_{i j} + ... \ , \num$$
where $R_{i j}$ is the Ricci tensor obtained using the ``seed'' metric 
$h_{i j}$, $R = h^{i j} R_{i j}$, $D_i$ is the covariant derivative 
compatible with the ``seed'' $h_{i j}$,
$${}^{(0)}\gamma_{i j} = A_0(t) h_{i j} \ , \num$$ 
$${}^{(2)}\gamma_{i j} = A_0(t) \left[f_2(t) R_{i j}+ {1 \over 3}
\left(g_2(t) - f_2(t)\right) R h_{ij}\right] \ , \num$$
and
$${}^{(4)}\gamma_{i j} = A_0(t) \left[{1 \over 3}\left(a_4(t) - b_4(t)
\right) R^2 h_{i j} + b_4(t) R R_{ij} + {1 \over 3}\left(c_4(t) - d_4(t)
\right) R^{k l} R_{k l} h_{i j} +\right.$$
$$\left.~~~~~~~~~~~~~~~~~~d_4(t) R_j^k R_{k i} + {1 \over 3}\left(e_4(t) 
- f_4(t) - g_4(t)\right)\left(D_k D^k R\right) h_{i j} + f_4(t) D_i D_j R 
+ \right.$$
$$\left.g_4(t) D_k D^k R_{i j}\right] \ . \num$$
The higher order terms in the expansions above are ``small'' in the sense 
that $R_{i j}$ contains two spatial gradients.  That is, if $L$ is the 
characteristic scale on which the fields vary, then $R_{i j} \sim L^{-2} 
h_{i j}$, $D_i D_j R_{k l} \sim L^{-4} h_{i j} h_{k l}$, and so on.

When the $K = k(t)$ slicing condition is imposed, the lapse function is 
no longer freely specified, but is obtained from
$$N = - {1 \over k(t)} {\dot{\sqrt{\gamma}~} \over \sqrt{\gamma}}\ , 
\num$$
once $k(t)$ is given and the form of $\gamma$ is known.  The 3+1 field 
equations also reflect the difference, becoming
$$16 \pi \rho = \tilde{R} - K^i_j K^j_i + k^2(t) \ , \num$$
$$8 \pi {\cal J}_i = \tilde{D}_j K^j_i \ , \num$$
$$8 \pi {\cal S}_i^j = \tilde{R}_i^j + k(t) K_i^j + 
{k(t) \over \left({\rm ln} \sqrt{\gamma}\right)\dot{}} \dot{K}_i^j -  
{1 \over \left({\rm ln} \sqrt{\gamma}\right)\dot{}} \tilde{D}_i 
\tilde{D}^j \left({\rm ln} \sqrt{\gamma}\right)\dot{} - 4 \pi 
\left(\rho - {\cal S}\right) \delta_i^j \ , \num$$
and
$$\dot{\gamma}_{i j} = {2 \over k(t)} {\dot{\sqrt{\gamma}~} \over 
\sqrt{\gamma}} K_{i j} \ . \num$$
Determining $N$ from Eq. (38) is unconventional.  However, it will be 
shown to be equivalent to the conventional approach (which is to solve 
Eq. (8)) in the next section.

The ordinary differential equations that determine the time-dependent 
coefficients in Eqs. (32-34) come about by putting the expansions in Eqs.  
(32-34) into the 3+1 equations just above, and then rearranging until 
all ``like'' terms ({\it i.e.}, quantities containing $h_{ij}$, $R h_{ij}$, 
or $R_{ij}$, {\it etc}.) are gathered.  One then gets as many equations as 
coefficients, since at each order the tensors $R h_{ij}$, $R_{ij}$, 
{\it etc}., are in general linearly independent.  We will not list the 
zeroth-order equations and solutions here since they are the same as 
those presented in the last section (except that ${}^{(0)}K$ is to be 
replaced with $k(t)$).  The second-order equations for $f_2(t)$ and 
$g_2(t)$ are
$$\left[k^{2(\Gamma - 1)/\Gamma} f_2^{\prime} \right]^{\prime} =
- {8 \over \Gamma^2 A_0(k)} k^{- 2 (1 + \Gamma)/\Gamma} \num$$
and
$$k g_2^{\prime} = - {2 \left(2 - 3 \Gamma\right) \over \Gamma^2 k^2 
A_0(k)} \ , \num$$
where ${}^{\prime} = {\rm d}/{\rm d}k = \dot{k}^{- 1} {\rm d}/{\rm d}t$ 
and $A_0(k) \propto k^{- 4/3 \Gamma}$.  The fourth-order equations are 
$$k a_4^{\prime} = {2 \left(3 \Gamma - 2\right) \over 3 \Gamma^2 k^2 
A_0(k)} \left(f_2 - g_2\right) - {2 - \Gamma \over 8 A_0(k)} \left(k 
f_2^{\prime}\right)^2 - {k \over 3} \left(f_2 f_2^{\prime} - g_2 
g_2^{\prime}\right) - {\left(2 - 3 \Gamma\right)^2 \over \Gamma^3 k^4 
A^2_0(k)} \ , \num$$
$$\left[k^{2 (\Gamma - 1)/\Gamma} \Sigma_4^{\prime}\right]^{\prime} = 
{2 - 3 \Gamma \over \Gamma^2 A_0(k)} k^{- 2 (1 + \Gamma)/\Gamma} 
\left(k f_2^{\prime}\right) + {4 \over 3 \Gamma^2 A_0(k)} k^{- 2 (1 +
\Gamma)/\Gamma} \left(7 f_2 + 2 g_2\right) -$$
$${4 \left(2 - 3 \Gamma\right) \over \Gamma^3 A_0^2(k)} 
k^{2(1 + 2 \Gamma)/\Gamma} \ , \num$$
$$k c_4^{\prime} = f_2 \left(k f_2^{\prime}\right) +
{3 (2 - \Gamma) \over 8} \left(k f_2^{\prime}\right)^2 +  
{2 \left(2 - 3 \Gamma\right) \over \Gamma^2 k^2 A_0(k)} f_2 \ , \num$$
$$\left[k^{2(\Gamma - 1)/\Gamma}\left(d_4^{\prime} - f_2 
f_2^{\prime}\right)\right]^{\prime} = - {16 \over \Gamma^2 A_0(k)} 
k^{- 2 (1 + \Gamma)/\Gamma} f_2 \ , \num$$
$$k e_4^{\prime} = {3 \Gamma - 2 \over 3 \Gamma^2 k^2 A_0(k)} \left(f_2 
- 4 g_2\right) + {4 \left(3 \Gamma - 2\right) \over \Gamma^3 k^4 A^2_0(k)} 
\ , \num$$
$$\left[k^{2(\Gamma - 1)/\Gamma}f_4^{\prime}\right]^{\prime} = {4 \left(2 
- 3 \Gamma\right) \over \Gamma^3 A_0^2(k)} k^{- 2 (1 + 2 \Gamma)/\Gamma} 
- {4 \over 3 \Gamma^2 A_0(k)} k^{- 2 (1 + \Gamma)/\Gamma} \left(f_2 - 
g_2\right) \ , \num$$
and
$$\left[k^{2(\Gamma - 1)/\Gamma} g_4^{\prime}\right]^{\prime} = 
{4 \over \Gamma^2 A_0(k)} k^{- 2 (1 + \Gamma)/\Gamma} f_2 \ , \num$$
where 
$$\Sigma_4^{\prime} \equiv b_4^{\prime} + {\Gamma \over 4} k g_2^{\prime} 
f_2^{\prime} + {1 \over 3} \left[f_2 (f_2 - g_2)\right]^{\prime} \ . 
\num$$
The equations for the time dependent coefficients for $\rho$ and 
${\cal J}_i$ are obtained from Eqs. (70) and (71) in Appendix A. Also note 
that we have still not specified the explicit form for $k(t)$.  

It is not crucial that we write down here the general solutions to these 
equations ({\it i.e.}, the sum of the ``homogeneous'' and ``particular'' 
solutions), even though it is straightforward to integrate and determine 
their form [24].  The crucial thing is to notice that Eqs. (44), (45), 
(47), and (49) have only one derivative.  In the synchronous gauge 
approach of Ref. [1] the equations all have two derivatives.  Therefore, 
there are ``homogeneous'' solutions appearing in the synchronous gauge 
calculation that do not appear here.  For instance, the general solution 
for $g_2(k)$ is 
$$g_2(k) = \gamma_1 - {3 \over \Gamma} k^{2(2 - 3 \Gamma)/3 \Gamma} \ , 
\num$$
whereas the general solution of Ref. [1] is 
$$g_2(t) = \gamma_1 + \gamma_2 t^{- 1} - {1 \over 4} {9 \Gamma^2 \over 9 
\Gamma - 4} t^{- 2(2 - 3 \Gamma)/3 \Gamma} \ . \num$$

To compare these two we must remove the final arbitrariness left in our 
solutions by giving an explicit form for $k(t)$.  The appropriate choice 
is $k(t) \propto t^{- 1}$ since the zeroth-order solution for the 
extrinsic curvature obtained by Ref. [1] has a trace given by ${}^{(0)}K 
\propto t^{- 1}$.  More generally, since $A_0(k) \propto k^{- 4/3 \Gamma}$ 
then the trace of the extrinsic curvature must decay with time if the 
zeroth-order ``expansion factor'' $A_0$ is to grow with time.  A trace 
like $k(t) \propto t^{- 1}$ thus results in a power law time dependence 
for $A_0$.  It now follows that the time dependence of the ``particular'' 
solutions in Eqs. (53) and (54) is the same ($t^{- 2 (2 - 3 \Gamma)/3 
\Gamma}$).  The synchronous gauge result for $g_2$ has the extra 
``homogeneous'' solution $\gamma_2 t^{- 1}$.  However, Comer {\it et al.} 
[1] demonstrated a coordinate transformation within the synchronous gauge 
exists whereby this extra term can be removed.  The $K = k(t)$ slicing 
condition automatically removes the extra term.  It is clear that pure 
gauge terms will automatically be removed at the fourth-order as well.

Thus, as claimed in the Introduction, there are pure gauge terms appearing 
in the synchronous gauge approach that are not present here.  However, we 
{\it do} recover the other terms that Ref. [1] obtained.  And like the 
results of Ref. [1], it can be shown [24] that some of the ``homogeneous'' 
solutions (for instance, the $\gamma_1$ term in Eq. (53)), can either be 
absorbed into a redefinition of the ``seed'' or decay with time for all 
allowed values of $\Gamma$.  The other ``homogeneous'' solutions, as well 
as the ``particular'' solutions, decay if $\Gamma < 2/3$, that is, when 
the matter is `inflationary.'  As explained in Ref. [1], this means that 
the metric perturbations will all freeze out once the comoving scale of 
the Hubble radius (which is shrinking like $t^{1 - 2/3 \Gamma}$) becomes 
less than the characteristic scale $L$ associated with the perturbations.

When $\Gamma > 2/3$, then the opposite behaviour happens.  Instead of 
decaying with time, the metric perturbations will all thaw out, and hence 
grow, as they enter the Hubble radius.  In this case, one must keep more 
and more terms if the solutions are going to be accurate [23].  It is for 
precisely this reason that the $K = k(t)$ condition, and a non-zero shift 
vector, can be of great help.  The differential equations for the 
time-dependent coefficients are simpler, and it may be possible to make 
the expansions converge faster, in the sense that a fewer number of 
iterations would be required.
\bigskip
\noindent{{\bf 5. The Lapse Function Solution}}
\bigskip
In this section we will verify that the lapse function used in the 
previous section does indeed satisfy Eq. (8).  Recall that we found in 
Sec. 3 that the spatial gradients of $N$ had to be small.  With that in 
mind, $N$ is expanded like
$$N = N_0(t) + N_2(t) R + N_4(t) R^2 + \eta_4(t) R^j_i R^i_j + n_4(t) 
D_i D^i R + ... \ . \num$$
The equation that determines $N$ is the appropriate form of Eq. (8), 
setting $N^i = 0$ and using $K = k(t)$:
$$\tilde{D}^i \tilde{D}_i N - 4 \pi N \left(\rho + {\cal S}\right) - 
N K^j_i K^i_j = - \dot{k}(t) \ . \num$$
The zeroth-, second-, and fourth-order equations for $N$ are constructed 
in the same manner as the previous section.  (For the reader's convenience, 
the expansions to fourth-order for $\rho$ and $K^j_i K^i_j$ are listed in 
Appendix A.)  But unlike the previous section, the equations for $N_0(t)$, 
$N_2(t)$, {\it etc.}, are algebraic [24].  

The solution for $N$ is 
$$N = {2 \over \Gamma} {\dot{k} \over k^2} + {2 - 3 \Gamma \over \Gamma^2 
k^4 A_0(k)} \dot{k} R - \left[{2 - 3 \Gamma \over \Gamma^2 k^4 A_0(k)} 
f_2 + {3 \left(2 - \Gamma\right) \over 16 k^2} \left(k 
f_2^{\prime}\right)^2\right] \dot{k} R^j_i R^i_j +$$
$$\left[{2 - 3 \Gamma \over 3 \Gamma^2 k^4 A_0(k)} \left(f_2 - g_2\right) 
+ {\left(2 - 3 \Gamma\right)^2 \over 2 \Gamma^3 k^6 A_0^2(k)} + {2 - 
\Gamma \over 16 k^2} \left(k f_2^{\prime}\right)^2\right] \dot{k} R^2 +$$
$$\left[{2 - 3 \Gamma \over 6 \Gamma^2 k^4 A_0(k)} \left(f_2 - 4 g_2\right) 
+ {2 \left(2 - 3 \Gamma\right) \over \Gamma^3 k^6 A_0^2(k)}\right] \dot{k} 
D_i D^i R \ . \num$$
Note the requirement $N > 0$ implies $\dot{k}/k^2 > 0$.  Also notice that 
Eq. (57) does indeed agree with the result obtained from Eq. (38), making 
use of Eq. (67) from Appendix A as well as Eqs. (43-51) for the metric 
corrections.

The time-dependence of $N$ can be qualitatively understood by using again 
$k(t) \propto t^{-1}$ and then inserting $f_2 \sim g_2 \sim t^{- 2 (2 - 3 
\Gamma)/3 \Gamma}$.  It is seen that $N_2/N_0 \sim t^{- 2 (2 - 3 
\Gamma)/3 \Gamma}$ and $N_4/N_0 \sim \eta_4/N_0 \sim n_4/N_0 \sim t^{- 4 
(2 - 3 \Gamma)/3 \Gamma}$.  Thus the corrections to $N$ all decay with 
time when $\Gamma < 2/3$, {\it i.e.}, the matter is `inflationary.'  
Hence, there is no contradiction with simultaneously having a monotonic 
$k(t)$ and a positive lapse function.   On the other hand, the corrections 
all grow if $\Gamma > 2/3$.  Hence, one must be cautious in the use of the 
long wavelength solutions [23] since the corrections to $N$ are not 
positive-definite.  (The second-order term is positive at all times when 
$\Gamma > 2/3$ if the ``seed'' is such that $R < 0$.)  Certainly, $N$ is 
positive for the amount of time that the second- and fourth-order
corrections remain smaller than the zeroth-order term.  After that, the 
sign of $N$ will depend on the particular value for $\Gamma$ and the form 
of the ``seed.''       
\bigskip
\noindent{{\bf 6. Conclusion}}
\bigskip
The main point of the long wavelength iteration scheme is to provide 
analytical cosmological models that contain large scale inhomogeneities.  
These solutions are perturbative and are obtained as expansions in spatial 
gradients.  The fundamental difficulty with this scheme is the lack of an 
absolute notion of space, and thus an absolute notion of spatial gradient.  
Any such notion must be introduced by hand.  Therefore, it is essential to 
verify that these long wavelength inhomogeneities result from real physics, 
and not from embedding unnecessarily wrinkled spacelike slices into an 
otherwise homogeneous spacetime.  

We have addressed this issue by applying the long wavelength iteration 
scheme to the 3+1 form of the Einstein-Perfect Fluid field equations.  We 
have verified the results of Ref. [1] using the $K = k(t)$ slicing 
condition, and in the process demonstrated that the 3+1 formalism makes it 
easier to implement the long wavelength iteration scheme.  In particular, 
it was seen that a non-zero shift vector can be used to exponentially 
dampen the effect of anisotropy on the extrinsic curvature and also 
spurious gauge modes that arise in the synchronous gauge approach do not 
show up when the $K = k(t)$ slicing condition is invoked. 

Even though we have gone a long way in applying the 3+1 formalism to the 
long wavelength iteration scheme, there are still some questions that need 
to be considered.  For instance, we did not solve the 3+1 equations in the 
canonical way, {\it i.e.}, as an initial-value problem.  It is clear that 
this should be investigated, since ultimately the question of how many 
terms to keep in the expansions should boil down to conditions on the 
initial data set.  The problem [25] can be framed around York's procedure 
for handling the constraints [8].  That is, one performs a conformal 
transformation on the metric, extrinsic curvature, and matter variables, 
and then uses York's covariant decomposition of symmetric tensors into 
their trace-free, vector, and trace parts.  There is even a small clue in 
what has been presented here that this is an appropriate way to proceed: 
the three-metric naturally had a conformal factor appear, which was the 
third-root of the determinant of the three-metric.  This is precisely the 
factor suggested by York's procedure.

Finally, it would also be interesting to see how to construct and use 
``minimal-strain'' or ``minimal-distortion'' shift vectors [10] in the 
long wavelength iteration scheme.  Smarr and York [10] show that these 
vectors are very adept at simplifying the form of the three-metric, by 
separating as completely as possible the purely ``kinematical'' (in the 
words of Ref. [10]) from the dynamical parts of the three-metric.  This 
is presently being investigated [25].  If time-independent 
``minimal-strain'' or ``minimal-distortion'' shift vectors can be 
constructed, then they can be immediately placed into Eq. (28). 
\bigskip
\noindent{{\bf Acknowledgements}}
\bigskip
I thank Tsvi Piran for suggesting the use of $K = k(t)$ hypersurfaces.  I 
also thank Nathalie Deruelle and Frank Estabrook for useful comments. 
Finally, I gratefully acknowledge partial support from a 1996 NASA/ASEE 
Summer Faculty Fellowship with the Jet Propulsion Laboratory of the 
California Institute of Technology.
\bigskip
\noindent{{\bf References}}
\bigskip
{\parindent 0pt
\baselineskip=18pt
[1] G. L. Comer, N. Deruelle, D. Langlois, and J. Parry, Phys. Rev. D 
{\bf 49}, 2759 (1994).

[2] N. Deruelle and Dalia S. Goldwirth, Phys. Rev. D {\bf 51}, 1563 
(1995).

[3] N. Deruelle and D. Langlois, Phys. Rev. D {\bf 52}, 2007 (1995).

[4] E. M. Lifschitz and I. M. Khalatnikov, Adv. Phys. {\bf 12}, 185 (1963).

[5] D. S. Salopek, Phys. Rev. D {\bf 43} 3214 (1991); D. S. Salopek and 
J. M. Stewart, Phys. Rev. D {\bf 47} , 3235 (1993); J. Parry, D. S. 
Salopek, and J. M. Stewart, Phys. Rev. D. {\bf 49}, 2872 (1994); and K. M. 
Croudace, J. Parry, D. S. Salopek, and J. M. Stewart, Ap. J. {\bf 423}, 22 
(1994).

[6] Jiro Soda, Hideki Ishihara, and Osamu Iguchi, Prog. Theor. Phys. 
{\bf 94}, 781 (1995).

[7] G. L. Comer, Nathalie Deruelle, and David Langlois, preprint 
gr-qc/9605042.

[8] J. W. York, ``Kinematics and dynamics of general relativity,'' in 
{\it Sources of Gravitational Radiation} (Cambridge University Press, 
Cambridge, 1979), ed. L. Smarr, pp. 83-126.

[9] Tsvi Piran, ``Methods of numerical relativity,'' in {\it Gravitational 
Radiation} (North-Holland Publishing Company, Amsterdam, 1982), eds. 
Nathalie Deruelle and Tsvi Piran, pp. 203-256.

[10] Larry Smarr and James W. York, Jr., Phys. Rev. D{\bf 17}, 2529 
(1978).

[11] S. W. Hawking and G. F. R. Ellis, {\it The Large Scale Structure of 
Space-Time} (Cambridge University Press, New York, 1973), pp. 96-102. 

[12] J. W. York, Phys. Rev. Lett. {\bf 28}, 1082 (1972).

[13] Tsvi Piran, J. Comp. Phys. {\bf 35}, 254 (1980).

[14] The $K = k(t)$ slicing condition also causes the Hamiltonian and 
momentum constraints ({\it cf.}, Eqs. (4) and (5)) to be completely 
decoupled (see Ref. [10]).

[15] A. Lichnerowicz, J. Math. Pure Appl. {\bf 23}, 37 (1944).

[16] M. P. Ryan, Jr. and L. C. Shepley, {\it Homogeneous Relativistic 
Cosmologies} (Princeton University Press, Princeton, 1975).

[17] M. Cantor, A. Fischer, J. Marsden, N. \'O Murchadha, and J. W. York, 
Commun. Math. Phys. {\bf 49}, 187 (1976).

[18] A. J. Goddard, Commun. Math. Phys. {\bf 54}, 279 (1977).

[19] Dalia S. Goldwirth and Tsvi Piran, Phys. Rev. D {\bf 40}, 3263 
(1989).

[20] K. Tomita, Prog. Theor. Phys. {\bf 48}, 1503 (1972); Prog. Theor. 
Phys. {\bf 54}, 185 (1975).

[21] C. W. Misner, K. S. Thorne, and J. A. Wheeler, {\it Gravitation}
(Freeman Press, San Francisco, 1973).

[22] Our definition for the extrinsic curvature agrees with that of Ref. 
[8] but differs by a minus sign and a factor of two from that of Ref. [1].

[23] As discussed in Ref. [1], one can think of an expansion like 
${\rm exp} x = 1 + x + x^2/2! + ...~$.  When $x \ll 1$ then the extra 
terms make only refinements, but when $x \geq 1$ they are a necessity.

[24] G. L. Comer, unpublished notes (1996).

[25] G. L. Comer, in preparation (1996).}

\bigskip
\noindent{{\bf Appendix A}}
\bigskip
For the convenience of the reader we list here some basic formulas used to 
build the second- and fourth-order equations given in the main text.  In 
all that follows consider that the three-metric has the form
$$\gamma_{i j} = {}^{(0)}\gamma_{i j} + \delta \gamma_{i j} \ , \num$$
where 
$$\delta \gamma_{i j} = {}^{(2)}\gamma_{i j} + {}^{(4)}\gamma_{i j} \ , 
\num$$
and ${}^{(2)}\gamma_{i j}$ and ${}^{(4)}\gamma_{i j}$ are the same as Eqs. 
(36) and (37), respectively, in the main text.  

The first formula is that for the inverse metric $\gamma^{i j}$.  It is 
given by
$$\gamma^{i j} = {}^{(0)}\gamma^{i j} + \delta \gamma^{i j} \num$$ 
where
$$\delta \gamma^{i j} = - {}^{(0)}\gamma^{i k}~{}^{(0)}\gamma^{j l} 
\delta \gamma_{k l} + {}^{(0)}\gamma^{i l}~{}^{(0)}\gamma^{j m}~
{}^{(0)}\gamma^{k n}~{}^{(2)}\gamma_{m n}~{}^{(2)}\gamma_{k l} \ . \num$$
The inverse metric is necessary for constructing the connection 
coefficients, the Ricci tensor and scalar, and the extrinsic curvature.

Letting  
$${}^{(0)}\Gamma^{i}_{j k} = {1 \over 2}~{}^{(0)}\gamma^{i l} \left[
{}^{(0)}\gamma_{j l, k} + {}^{(0)}\gamma_{k l, j} - {}^{(0)}
\gamma_{j k, l}\right] \ , \num$$
then it can be shown [21] that the connection coefficients are
$$\tilde{\Gamma}^{i}_{j k} = {}^{(0)}\Gamma^{i}_{j k} + \delta 
\Gamma^{i}_{j k} \ , \num$$
where
$$\delta \Gamma^{i}_{j k} = {1 \over 2} {}^{(0)}\gamma^{i l} 
\left[{}^{(0)}D_k \delta \gamma_{j l} + {}^{(0)}D_j \delta \gamma_{k l} - 
{}^{(0)}D_l \delta \gamma_{j k} \right] \num$$
and ${}^{(0)}D_i$ is the covariant derivative compatible with
${}^{(0)}\gamma_{i j}$ ({\it i.e.}, ${}^{(0)}D_i~{}^{(0)}\gamma_{j k} = 
0$).

The Ricci tensor is obtained from
$$\tilde{R}_{i j} = {}^{(0)}\tilde{R}_{i j} + {}^{(0)}D_k \delta 
\Gamma^{k}_{i j} - {}^{(0)}D_j \delta \Gamma^{k}_{i k} \ . \num$$
In terms of the metric corrections given in Eqs. (44) and (45) then
$$\tilde{R}^j_i = {1 \over A_0} R^j_i + {1 \over A_0} \left[{1 \over 2} 
g_2 \left(D_i D^j R - {1 \over 3} D_k D^k R \delta^j_i\right) - 
{1 \over 2} f_2 D_k D^k\left(R^j_i - {1 \over 3} R \delta^j_i\right) 
+\right.$$
$$\left.3 f_2 \left(R^j_k R^k_i - {1 \over 3} R^k_l R^l_k \delta^j_i
\right) - {3 \over 2} f_2 R \left(R^j_i - {1 \over 3} R \delta^j_i\right) 
+ {1 \over 6} \left(f_2 - 4 g_2\right) D_i D^j R - f_2 R^j_k R^k_i 
+\right.$$
$$\left.{1 \over 3} \left(f _2 - g_2\right) R R^j_i\right] \ . \num$$

Derivatives of the determinant of the three-metric are obtained from 
$\delta \gamma = \gamma \gamma^{i j} \delta \gamma_{i j}$.  Letting 
$\delta$ be $\partial/\partial t$, then 
$${\dot{\sqrt{\gamma}~} \over \sqrt{\gamma}} = 3 H_0 + {1 \over 2} 
\dot{g}_2 R + {1 \over 2} \left(\dot{c}_4 - f_2 \dot{f}_2\right) R^k_l 
R^l_k + {1 \over 2} \left(\dot{a}_4 + {1 \over 3} \left[f_2 \dot{f}_2 - 
g_2 \dot{g}_2\right]\right) R^2 +$$
$${1 \over 2} \dot{e}_4 D_k D^k R \ , \num$$
where $H_0 = \dot{A}_0(t)/2 A_0(t)$.  Notice that it is only $g_2(t)$ 
which enters as a second-order term.

The extrinsic curvature is
$$K^j_i = {1 \over 3} k(t) \delta^j_i + {k(t) \over 6 H_0} 
\left[\dot{f}_2 \left(R^j_i - {1 \over 3} R \delta^j_i\right) + 
\left(\dot{b}_4 - {1 \over 6 H_0} \dot{f}_2 \dot{g}_2 + {1 \over 3} 
\left[f_2\left(f_2 - g_2\right)\right]{}^{\dot{}}\right) \times \right.$$
$$\left.R \left(R^j_i - {1 \over 3} R \delta^j_i\right) + \left(\dot{d}_4 
- f_2 \dot{f}_2\right) \left(R^j_k R^k_i - {1 \over 3} R^k_l R^l_k 
\delta^j_i\right) + \dot{f}_4 \left(D_i D^j R - {1 \over 3} D_k D^k R
\right) +\right.$$
$$\left.\dot{g}_4 D_k D^k\left(R^j_i - {1 \over 3} R \delta^j_i\right)
\right] \ . \num$$
Notice that the correcting pieces are such that we still maintain $K = 
K^i_i = k(t)$.  It is also found that
$$K^k_l K^l_k = {1 \over 3} k^2(t) + {k^2(t) \over (6H_0)^2} \dot{f}_2^2 
\left(R^k_l R^l_k - {1 \over 3} R^2\right) \ . \num$$
This has corrections beginning with the fourth-order.  In the synchronous 
gauge, there are corrections starting with the second-order.  However, it 
is really because $K = k(t)$ holds at {\it all} orders that our second- 
and fourth-order equations are simpler than those of the synchronous 
gauge.

Finally we list the equations---good to fourth-order---for $\rho$ and 
${\cal J}_i$:
$$16 \pi \rho = {2 \over 3} k^2(t) + {1 \over A_0} R + {1 \over 6 A_0} 
\left[f_2 - 4 g_2\right] D_i D^i R - \left[{1 \over A_0} f_2 + 
{k^2(t) \over (6 H_0)^2} \dot{f}_2^2\right] R^j_i R^i_j +$$
$${1 \over 3}\left[{1 \over A_0} \left(f_2 - g_2\right) + {k^2(t) \over 
(6 H_0)^2} \dot{f}_2^2\right] R^2 \num$$
and
$${48 \pi H_0 \over k(t)} {\cal J}_i = {1 \over 6} \dot{f}_2 D_i R + 
{1 \over 6} \left[3 \dot{g}_4 - \dot{b}_4 + {1 \over 6 H_0} \dot{f}_2 
\dot{g}_2 - {1 \over 3} \left(f_2 \left[f_2 - g_2\right]\right)^{\dot{}}
\right] R D_i R +$$
$$\left[\dot{b}_4 + {1 \over 3} \dot{f}_4 - {1 \over 2} \dot{g}_4 + 
{1 \over 2} \left(\dot{d}_4 - f_2 \dot{f}_2\right) - {1 \over 6 H_0} 
\dot{f}_2 \dot{g}_2 + {1 \over 3} \left(f_2 \left[f_2 - g_2\right]
\right)^{\dot{}}\right] R^j_i D_j R +$$
$$\left[\dot{d}_4 - f_2 \dot{f}_2 + 4 \dot{g}_4\right] R^j_k D_j R^k_i - 
\left[3 \dot{g}_4 + {2 \over 3} \left(\dot{d}_4 - f_2 \dot{f}_2\right)
\right] R^j_k D_i R^k_j +$$
$${1 \over 3} \left[2 \dot{f}_4 + {1 \over 2} \dot{g}_4\right] D_k D^k 
D_i R\ . \num$$
From this we see ${\cal J}_i {\cal J}_j$ has six spatial gradients.  
Hence, $\tilde{u}_i \tilde{u}_j$ can be ignored in all equations.  In 
particular, $\rho \approx \rho^*$ and ${\cal S}^j_i \approx \left(\Gamma 
- 1\right) \rho \delta^j_i$.
\bigskip
\noindent{{\bf Appendix B}} 
\bigskip
In this appendix we show how to solve Eq. (23) for ${}^{(0)}{\cal K}^j_i$.  
Let
$${\cal M}^{j k}_{~~~i l} \equiv \delta^j_l \tilde{D}_i N^k - \delta^k_i
\tilde{D}_l N^j \num$$
and $E^{a i}_{~~~j}$ represent `$a$' eigenvectors with eigenvalues 
$\lambda_{(a)}$, $a = 1, 2, 3, ... , 9$; that is,
$${\cal M}^{j k}_{~~~i l} E^{a i}_{~~~j} = \lambda_{(a)} E^{a i}_{~~~j} 
\ , \num$$
with no sum being done over $a$.  Also let $\left(E^{- 1}\right)
{}^{~~~j}_{a i}$ be such that 
$$E^{a i}_{~~~j} \left(E^{- 1}\right){}^{~~~j}_{b i} = \delta^a_b 
\qquad {\rm and} \qquad \sum_a  E^{a j}_{~~~i} \left(E^{- 1}\right)
{}^{~~~k}_{a l} = \delta^k_i \delta^j_l \ . \num$$
Then, the matrix with components ${\cal M}^{j k}_{~~~i l}$ can be 
diagonlized via
$$E^{a i}_{~~~j} {\cal M}^{j k}_{~~~i l} \left(E^{- 1}\right)
{}^{~~~k}_{b l} = \lambda_{(a)} \delta^a_b \ . \num$$
Finally, let 
$${\cal K}^a \equiv E^{a i}_{~~~j} {}^{(0)}{\cal K}^j_i \qquad {\rm and} 
\qquad {}^{(0)}{\cal K}^j_i \equiv \sum_a \left(E^{- 1}\right)
{}^{~~~j}_{a i} {\cal K}^a \ . \num$$
Therefore, Eq. (23) becomes
$$\dot{\cal K}^a - \left(\tilde{D}_k N^k\right) {\cal K}^a - \lambda_{(a)} 
{\cal K}^a = 0 \num$$
and the solutions are thus
$${\cal K}^a = {\rm exp}{\left[\left(D_k N^k + \lambda_{(a)}\right) t
\right]}~{\cal C}^a \ , \num$$
with ${\cal C}^a$ only depending on $x^i$.  Or, letting
$${\cal G}^j_i \equiv \sum_a {\rm exp}{\left(\lambda_{(a)} t\right)} 
~\left(E^{- 1}\right){}^{~~~j}_{a i}{\cal C}^a \qquad , \qquad 
{\cal G}^j_j = 0 \ , \num$$
we get the solution written in Eq. (28).

\end